\documentclass{JHEP}

\def\a{\alpha}

\def\d{\delta}

\newcommand{\ep}{\epsilon}

\newcommand{\om}{\omega}
\newcommand{\Om}{\Omega}

\newcommand{\si}{\sigma}
\newcommand{\Si}{\Sigma}

\def\Th{\Theta}

\def\det{{\rm det}}

\def\Vol{{\rm Vol}}
\def\exp{{\rm exp}}

\def\llap#1{\hbox to 0pt{\hss#1}}
\def\pola{a\llap{\hbox{\char'30\kern-1.2pt}}}
\def\pole{e\llap{\hbox{\char'30\kern-.8pt}}}

\newcommand{\vs}{\vskip.3cm}
\newcommand{\non}{\nonumber\\}

\newcommand{\twomat}[4]{\left( \begin{array}{cc} #1&#2\\ 
#3&#4\end{array}\right)}
\newcommand{\p}{\partial}

\def\half{{\mbox{\small  $\frac{1}{2}$}}}

\newcommand{\lb}{\left( }
\newcommand{\rb}{\right) }

\newcommand{\beq}{\begin{equation}}
\newcommand{\eeq}{\end{equation}}
\newcommand{\beqa}{\begin{eqnarray}}
\newcommand{\eeqa}{\end{eqnarray}}
\newcommand{\ben}{\begin{enumerate}}
\newcommand{\een}{\end{enumerate}}
\newcommand{\bit}{\begin{itemize}}
\newcommand{\eit}{\end{itemize}}

\newcommand{\refeq}[1]{(\ref{#1})}

\def\Xh{{{\widehat  X}}}
\def\Sie{{\widehat \Si}}
\def\adss{$AdS_3\times S^3\;${}}

\def\ninfty{$Q_5\to\infty\;${}}

\title{SU(2) WZW D-branes\\ and
their noncommutative geometry\\
\vs
from DBI action
\thanks{Work supported in part
by Polish State Committee for Scientific Research (KBN) under contract
2P 03 B03 715 (1998-2000) and the Alexander-von-Humboldt Foundation.}
}
\author{      
Jacek Pawe\l czyk
\\
Sektion Physik, Universit\"at M\"unchen\\
Theresienstrasse 37
80333 M\"unchen, Germany\\
and\\
{\sl Institute of Theoretical Physics\\ Warsaw University, Ho\.{z}a 69,
PL-00-681 Warsaw, Poland}\\ \email{Jacek.Pawelczyk@fuw.edu.pl}
}
\abstract{
 Using properties of the DBI action we find 
D-branes on $S^3$ of the radius $Q_5$ corresponding to the conjugacy
classes of SU(2). The branes are stable due to nonzero
2-form NSNS background. 
In the limit of large $Q_5$ the dynamics of branes
is governed by the non-commutative Yang-Mills theory.  The results
partially overlap with those obtained in the recent paper hep-th/0003037.
}

\begin{document}
Recently it has been discovered that in a special limit the dynamics
of the matrix  model is described by the non-commutative
Yang-Mills theory \cite{douglas}. This has sparked new interest in 
NCG for strings propagating in NSNS antisymmetric tensor field
background \cite{others}.
Introduction of D-branes  has provided deeper
understanding of the role of non-commutativity \cite{s-w-nc} and 
it has allowed to derive 
conditions under which NCG starts to play dominant role in the dynamics of
strings. It has also led to new  understanding of the 
connection between quantum groups and WZW models
\cite{ale-schum,ale-schum-reck}.  These two papers have used the standard CFT
language  
what is a drastic bound on possible applications. In particular it does not
allow to 
analyze RR backgrounds so much studied in the context of Maldacena's
conjecture \cite{maldacena}.  

The purpose of this paper is
to provide understanding of the results of \cite{ale-schum,ale-schum-reck}
in the more universal language then that of WZW models. The hope is that after
taking this lesson one would be able to derive interesting results for more
general string/M-theory backgrounds. 
Thus we shall describe various branes on the background of
SU(2) WZW model using the D-brane effective action (DBI action) only.
We shall also
show how the  non-commutativity  appears in this approach.
Methods applied here are limited to the case of large level of the SU(2) WZW
model 
 what in gravity language means large radius of the $S^3$.

Let us recall some of the results of \cite{ale-schum} and
\cite{ale-schum-reck}. 
D-branes in the level $k$ SU(2) WZW model are in one-to-one
correspondence with 
special integer conjugacy classes $ghg^{-1}$ for some fixed $h$
\cite{ale-schum}. There are  
 $k+1$ of them: two D-particles ($h=\pm 1$) and $k-1$ D2-branes corresponding
to two-spheres. The n-th sphere passes through the point 
$\exp(i\pi n\si^3/k)\in\,$SU(2), $n=1...k-1$. We must also stress that
D3-branes and 
D1-branes are excluded from this list.
For large $k$ the 2-spheres are in fact so-called fuzzy spheres \cite{johnm}.

The example of string theory background which involves the level $k$ SU(2) WZW
model is the near horizon limit of the 
F1, NS5 system (see e.q. \cite{mald-strom}). Below we write only the relevant
terms  
\beqa
ds^2/\a' &=&Q_5\; d\Omega_3^2\non
H^{NSNS}/\a'&=& 2Q_5\;\ep_3\non
e^{2\phi }& =& const.\label{back}
\eeqa
where $Q_5$ denotes the number of NS5-branes and it is equal to the level $k$
of 
the SU(2) WZW model, $\ep_3$ is the volume element of the 
unit 3-sphere.

The effective action of the D-branes is given by DBI expression
\beq\label{dbi}
S_{DBI}=-T_p\int_{\Vol}
  e^{-\phi}\sqrt{-\det[(X^*G+2\pi\a'F+X^*B)_{ab}]}
\eeq 
In the following we shall discuss  classical configurations of branes embedded
in $S^3$ of \refeq{back}.
Before we start to analyze equation of motion resulting from \refeq{dbi} 
we state several assumptions we make which seems to be
natural here.
We require the string coupling constant to be small and $Q_5$ to be large in
which case \refeq{back} is the part of an  exact string 
background as 
at this limit supergravity is the perfect description of string theory. 
Moreover D-branes can 
be described completely classically by the DBI action. 
We also assume that the higher order correction to the DBI action are
negligible. 
We shall be interested here
only in the $S^3$ part of the configuration thus it is even irrelevant if we
consider IIB (as above) or IIA string. Thus some of the arguments given in our
paper could be easily 
generalized to branes  embedded in a 10d manifold of
the form $S^3\times M^7$ under the condition that the 
 embeddings are of the product structure i.e. 
the induced metric, pull-back of $B$ and $F$ fields are of block diagonal form.

Recall that  D-branes are defined to be the ends of the open strings.
The string couple to the external sources (gauge $A$ and $B$ field) as follows 
\beq\label{coup}
\exp[{\frac{i}{2\pi\a'}(\int_{\p\Si}2\pi\a'X^*A+\int_\Si X^*B)}]
\eeq
The example of the WZW model shows that the above formula can not be well
defined 
globally for topologically non-trivial $A$ and $B$ fields. It is known that
for closed strings $\p\Si=0$ the proper formula is 
$\exp[\frac{i}{2\pi\a'}\int_{\Sie} \Xh^*H)]$ for $H$ being locally $dB$ and
$\Xh$ is an extension of $X$ to a 3-manifold $\Sie$ such that 
$\p \Sie=\Si$. Now we consider configuration of D-brane embedded into
submanifold $M_D$ of the target space manifold $M$.
One must repeat the above construction for the open string case
\cite{ale-schum,gawedzki}. For the world sheet with one boundary the
appropriate 
3-manifold must respect $\p \Sie=\Si+D^2$.
Rewriting the WZW model with boundary we get the proper global form of
\refeq{coup} \footnote{$-$ in front of the first term is due to different
  orientation of 
  boundary $-\p \Si=\p D^2$ in $D^2$ compare to $\Si$.}
\beq\label{unb-int}
\exp\left[\frac{i}{2\pi\a'}\left(-\int_{D^2}\Xh^*(2\pi\a' F+B)+ 
\int_\Sie\Xh^*H\right)\right]
\eeq
where $\Xh$ is an extension of 
$X(\p\Si)$ to a full 2-disc $D^2$ such that $\Xh(D^2)\subset M_D$ ($F\neq 0$
only on the D-brane manifold $M_D$). 
We stress that \refeq{unb-int} has proper gauge invariance and for
topologically trivial $H$ it reduces to 
\beq
\exp\left[\frac{i}{2\pi\a'}\left(-\int_{D^2}\Xh^*(2\pi\a' F)+
\int_\Si X^*B\right)\right]
\eeq
i.e. to \refeq{coup}. 
Notice that one must be able to
define $B$ on any $\Xh(D^2)$ 
thus we must have  $[H]_{M_D}=0$.
The value of the integral \refeq{unb-int} should not
depend on the way one make the extension. 
This forces to put
\beqa
\frac{i}{2\pi\a'}\left[\int_{C^2}(2\pi\a' F+B)
-\frac{i}{2\pi\a'}\int_{C^3} H \right]= 2i\pi m
\eeqa
where $C^2\in H_2(M_D)$ and $C^3\in H_3(M,M_D)$.{} Here a note is necessary
concerning topology of the problem. For the argument we need an exact sequence
of homologies 
\beq
\dots\to H_3(M_D)\to H_3(M)\to H_3(M,M_D)\to H_2(M_D)\to H_2(M)\to\dots
\eeq
If we assume that $H_3(M_D)=H_2(M)=0$ then all cycles of $H_3(M)$ are in
$H_3(M,M_D)$. 
Then one can write $\int_{C^3}H=\int_{C^2}B$ mod 
$\frac{i}{2\pi }\int_{C^3_M}H=2\pi Q_5 m$ for 
$C^3_M\in H_3(M)$. Thus the quantization condition reads
\beq\label{first}
\int_{C^2}F=2 \pi n, \quad n\in \mbox{ \bf Z}
\eeq
and $n$ is defined modulo $Q_5$. This is the same as postulated in
\cite{bds}. 
 In the above we have disregarded the difference between the cycles 
in $M$ and $M_D$ and their image given by $\Xh$.

{\bf D3 brane.} Here we discuss the Dp-branes wrapped on the entire $S^3$. 
According to 
the condition $[H]_{M_D}=0$ we see that such a wrapping is impossible. We
would like to provider here a different argument based on DBI action. 
First one must notice  
that due to $[H]_{M_D}=0$  the DBI action \refeq{dbi} is not well defined as 
 $B$ is not well defined on $S^3$. 

In order to be more specific we concentrate on D3 brane in the background 
\refeq{back} and change the brane description
to the dual form  of the DBI action discussed e.g. in \cite{dualDBI}.
It has the same classical solutions as \refeq{dbi} what is
the property we are interested in. 
\beq\label{ddbi}
S_{DBI}\propto \int_{\Vol}
  e^{\phi}\sqrt{-\det[(X^*G+2\pi\a'{\tilde F})_{ab}]}-\pi\a'{\tilde
F}\wedge B
\eeq
The last term come form CS part of the DBI action. Integrating it by parts we
get
\beq
\pi\a'{\tilde A}\wedge H^{NSNS}
\eeq
thus the action contains only the well defined $B$ field strength. 
With the $ H^{NSNS}$ background given by \refeq{back} we see that there is a
U(1) charge generated on the D3 world-volume. The charge can not stay on $S^3$
as it is a compact space, thus it forces the brane to partially unwrap the
sphere.  In the case of \adss the brane runs into the boundary of the AdS
space. The above argument follows the baryon construction of \cite{w-bar}. 

{\bf D2 brane.}
Here we concentrate upon D2-brane case totally wrapped on $S^3$.
It can be also a e.g. partially wrapped D3 brane.
We analyze its equation of motion and
find that contrary to the naive expectation the static brane it stable. As the
indication of stability we invoke the lack of the tachyonic mode for the
fluctuation of the brane.

In order to analyze the classical equations of motion we must find out the
pull back of $B$ field to the brane world-volume. On any 2-d submanifold of
$S^3$ the 
 $B$ filed is well (but not uniquely) defined. We have the freedom of changing
$B$ by an exact 2-form - in our case this will realized by choice
of the solution for $F_{12}$.
In the coordinates in which the
metric on $S^3$ is $ds^2=Q_5[d\phi^2+\sin^2(\phi) d\Om_2^2]$
we have for the chart covering $\phi=0$
\beq
B=Q_5\a'(\phi-\nu-\half\sin(2\phi))\,\ep_2
\eeq
where $\ep_2$ is the volume form of the unit $S^2$.

 We shall find  extrema of the DBI action corresponding to branes wrapped
on $S^2$ given by some constant angle $\phi$. 
 The Euler-Lagrange equations are respected by
\beq\label{fsol}
2\pi\a' F=-Q_5\a'(\phi-\nu)\,\ep_2 ,\quad \phi(x)=\phi=const.
\eeq
with all the other components of $F$ equals zero. 
We also set $\nu=0$ requiring that the charge and the tension of the 
$\phi=0$ brane be zero.
It is worth to note that the classical solution exists for  any angle $\phi$.
When we apply the quantization condition (\ref{first}) we get
\beq\label{fquant}
Q_5 \phi\,=2\pi n
\eeq
We remind that $n$ is defined only modulo $Q_5$.

We can compare \refeq{fquant} with results one gets 
assuming that the brane couple
to some RR fields i.e. carry RR charge   
\beq
+T_p\int  e^{2\pi\a'F+X^*B}\wedge\bigoplus_q C_q
\eeq
where, $T_{Dp}=1/({(2\pi)^p\a'^{(p+1)/2}g_s})$. 
The background $2\pi\a'F+X^*B$
generates RR charge of the D(p-2)-brane
equals to 
$$
T_p\int_{S^2} (2\pi\a'F+X^*B)= -T_p (4\pi\a' Q_5)\half\sin(2\phi).
$$
One expects that this charge is integer multiple of $T_{(p-2)}$
i.e.
\beq\label{quant}
\frac{1}{2\pi\a'}\int_{S^2}(2\pi\a'F+X^*B)=-2\pi\,n
\eeq
but this is in contradiction with \refeq{fquant} for finite $Q_5$. If one takes 
the $Q_5\to \infty$ limit then both formulae agree. \footnote{The gap between
  \refeq{fquant} and \refeq{quant} has been filled recently in \cite{taylor}.}

The second derivative of 
the DBI action with respect to the gauge fields and $\phi(x)$ gives kinematics of
fluctuations. One easily finds that fluctuations of $\phi(x)$ only are massive
but $\phi(x)$ mixes with gauge field $F$ leading to some massless
modes \cite{bds}. Thus there
is no tachyon in the spectrum and the brane configuration is stable.

{\bf Non-commutative geometry.} 
We can also claim that at the \ninfty limit some of the branes are described
by the non-commutative geometry. Here we follow the route of
\cite{s-w-nc}. First we notice that at the \ninfty limit we have
\beqa 
ds^2&=&\a'\frac{(n\pi)^2}{Q_5} d\Om_2^2\;\to 0\non
2\pi\a'F+X^*B&=&-\a' (n\pi) \,\ep_2
\eeqa
Thus the closed string metric goes to zero while induced $2\pi\a'F+X^*B$ is
constant on the D2-branes world-volume what is a good sign for the
non-commutativity. 
Next  we calculate the open string metric and the Poisson
structure inverting $2\pi\a'F+ X^*(B+g)$. The inverse matrix is
\beq\label{o-metric}
\lb\frac{1}{X^*(g+B)+2\pi\a' F}\rb=\frac{1}{Q_5 \sin\phi\a'}
\twomat{\sin\phi}{\cos\phi}{-\cos\phi}{\sin\phi} 
\eeq
Inverse of its symmetric part is the open string metric. We have 
$G_{ab}=\a'Q_5\d_{ab}$. Hence from the open string point of view all
 spheres have the same area! This, of course, is directly related
to the flat direction $\phi=const$ in the solution \refeq{fsol}.
The Poisson structure on $S^2$ (also called
deformation parameter) is
\beq
\Th^{12}=\frac{2\pi}{Q_5}\cot\phi \to \frac2{n}
\eeq
The symplectic structure is the inverse of the Poisson structure and it is
$\om_{12}=(n/2)$. One can check that this parameter precisely corresponds
to the symplectic 
structure used by Berezin in order to quantize
$S^2$ \cite{berezin}. The non-commutative version on this $S^2$ is called
the fuzzy spheres \cite{johnm}. From \cite{matrix} one may claim that
the Y-M theory on this sphere is a theory of $(n+1)\times(n+1)$
hermitian matrices. 
Such a Y-M theory has $(n+1)^2$ degrees of freedom. Here we must stress that 
these results are in full agreement with  \cite{ale-schum-reck}. It would be
interesting to make explicit comparison of the brane dynamics and the above
matrix model.

We conclude that the branes
dynamics is described by the non-commutative Y-M theory.
The branes world-volume are
2-spheres which are non-commutative manifolds  with the 
non-commutativity parameter $\Th^{12}=\frac2{n}$. \\

{\bf A note added.} Some of the results of this paper have been independently
obtained in the recent paper  \cite{bds}.

\acknowledgments{  
I am grateful S.Theisen for illuminating discussions and reading
the manuscript.
I also thank  K.Gaw{\c e}dzki, A.Alekseev, G. Arutyunov and A. Recknagel
for comments and interest in this work.

\def\thebib{}
\thebib

\end{document}